# First-principles-based search for emergent topological spin textures in transition-metal dichalcogenide monolayers

**Yuhyun Cha[1], Do Hoon Kiem[1,2] & Myung Joon Han[1]** ✉

Topological spin textures such as skyrmions and bimerons offer rich physical functionalities and strong potential for future spin-based technologies. Identifying materials that host such emergent magnetic states, however, remains challenging due to the complex interplay between microscopic magnetic interactions and large-scale spin textures. Here, using a multiscale first-principles discovery framework, we identify a broad class of two-dimensional materials that stabilize nontrivial spin textures. The framework integrates relativistic density functional theory, magnetic force theory, and atomistic spin-dynamics simulations, enabling a direct connection between electronic interactions and emergent magnetic textures and allowing efficient screening of systems with large magnetic unit cells and competing exchange interactions. Applying this approach, we systematically screen 312 monolayer transition-metal dichalcogenides across 1T and 2H phases of 3*d*, 4*d*, and 5*d* transition metals. We uncover a diverse set of emergent spin textures, including skyrmions, bimerons, and complex domain states, and clarify their distinct stabilization mechanisms driven by exchange interactions, Dzyaloshinskii-Moriya interactions, and magnetic frustration. These results identify promising candidate materials for topological spin textures and establish a transferable computational framework for accelerating materials discovery of complex magnetic states in two dimensions.

Magnetic domains and complex spin textures have emerged as a versatile platform for exploring novel physics and enabling new spintronic functionalities[1–4]. Structures such as skyrmions, moiré magnetism, and spin spirals exhibit a range of fascinating phenomena—including skyrmion Hall and topological Hall effects[5,6]. Their distinctive characteristics not only deepen our understanding of magnetic quantum matter but also hold promise for technological applications. In particular, topologically protected spin textures are regarded as promising building blocks for next-generation memory and logic devices. To realize such applications, however, it is essential to identify suitable materials that can host desired spin textures with high stability and controllability[7–10].

Among two-dimensional materials, monolayer transition-metal dichalcogenides (TMDCs) have attracted sustained attention owing to their structural versatility and tunable properties[11–16]. In particular, *Janus*-type TMDCs, where two different chalcogen species occupy the top and bottom layers, exhibit unique features such as enhanced Rashba spin-orbit coupling, out-of-plane piezoelectricity, and improved photovoltaic performance[17–21]. From a magnetic perspective, the broken inversion symmetry in these materials gives rise to intrinsic Dzyaloshinskii-Moriya (DM) interactions, which can stabilize various topological spin textures. Furthermore, the relative ease of thin-film synthesis and the tunability of magnetic properties through doping, strain, or layer thickness make them an attractive materials platform for exploring complex magnetic superstructures[22–30].

In parallel, high-throughput computational screening based on density functional theory (DFT) has recently become a powerful approach for accelerating materials discovery. Such efforts have achieved remarkable success in identifying candidates for energy, structural, and topological materials[31–45]. However, applying similar strategies to magnetic materials hosting complex spin textures has remained a formidable challenge. The intrinsic requirement of large magnetic unit cells, the need to incorporate spin-orbit coupling, and the multiplicity of nearly degenerate magnetic

[1]Department of Physics, Korea Advanced Institute of Science and Technology, Daejeon, Republic of Korea. [2]Materials Science and Technology Division, Oak Ridge National Laboratory, Oak Ridge, Tennessee 37831, USA. ✉e-mail: mj.han@kaist.ac.kr

states make straightforward high-throughput DFT screening practically infeasible.

To overcome these limitations, we develop an integrated multi-scale computational workflow that combines DFT, magnetic force response theory (MFT)[46–49], and atomistic spin simulations[50]. This framework connects microscopic electronic interactions with emergent spin textures in a computationally efficient manner, enabling large-scale magnetic material screening that was previously unfeasible. Using this workflow, we systematically investigate all possible 1T- and 2H-phase combinations of 3d, 4d, and 5d transition metals with S, Se, and Te—312 TMDC compounds in total. The reliability of the suggested workflow is verified by the calculated magnetic interaction profiles well respecting Moriya's symmetry rules[51].

Further, our results reveal a variety of topologically nontrivial spin textures, including: (i) skyrmions in 2H-RuSSe, 1T-FeSTe, 2H-TcSTe, and 1T-VSSe; (ii) bimerons in 1T-FeSeTe, 2H-$FeS_2$, 1T-$RuS_2$, and 1T-CrSeTe; and (iii) maze-like domain patterns (convertible into skyrmions under external fields) in 2H-FeSTe, 1T-MnSSe, 1T-$MnS_2$, and 1T-$MnSe_2$. In addition, we identify antiferromagnetic triangular systems exhibiting complex domain states such as Néel vortices and spirals. Detailed analyses elucidate the interplay between DM interactions and competing exchange couplings that underlie these textures. Overall, our findings not only uncover new candidates for topological spin textures in TMDCs but also establish a practical computational framework for multi-scale exploration of complex magnetic materials.

## Results

### Workflow overview

Figure 1a illustrates the proposed workflow, which consists of three main stages. The first stage involves DFT calculations. This step not only screens out non-magnetic or otherwise irrelevant materials, but also, for magnetic candidates, provides the spin-dependent Kohn–Sham Hamiltonian together with the optimized lattice geometry. These quantities serve as essential inputs for the subsequent stages of the workflow.

In this study, we consider transition metals from the 3*d*, 4*d*, and 5*d* series (Sc, Ti, V, Cr, Mn, Fe, Co, Ni, Cu, Y, Zr, Nb, Mo, Tc, Ru, Rh, Pd, Ag, Hf, Ta, W, Re, Os, Ir, Pt, and Au), combined with three chalcogens (S, Se, and Te). Both commonly observed structural phases of TMDCs, namely the 1T and 2H phases, are examined, and both homostructural and heterostructural (Janus) configurations are included (see Fig. S1a, b).

Figures 1b, c summarize the DFT results for a total of 312 TMDC compounds. The relative stability of the magnetic phases, as determined from DFT total energies, is indicated by the color of the upper-left triangular markers (see the red–blue color scale). A total of 258 materials are screened out because they fail to converge to a stable magnetic solution, as indicated by the cross symbols. The remaining 54 materials are then selected as candidates for the next stage of the workflow. Full numerical results—DFT total energies, magnetic moments, and lattice constants for all 312 compounds—are tabulated in Table S1.

The second step is to construct ab initio spin Hamiltonian through MFT calculations (see Fig. 1a). With the converged solutions of DFT, MFT estimates the magnetic interaction parameters, constructing the material-specific Hamiltonian:

$$\mathcal{H} = -\sum_{i,j} J_n^{ij}\mathbf{S}^i\cdot\mathbf{S}^j - \sum_{i,j}\mathbf{D}_n^{ij}\cdot(\mathbf{S}^i\times\mathbf{S}^j) - \sum_i K(\mathbf{S}^i\cdot\widehat{\mathbf{z}})^2. \quad (1)$$

Here $i$ and $j$ are the atomic site indices, and $\widehat{\mathbf{z}}$ denotes the unit vector pointing the anisotropy axis which is the out-of-plane direction in the current study. The lower index $n$ denotes the $n$-th neighboring interactions.

To determine the exchange components of the Hamiltonian, the MFT calculates interactions in momentum space via:

$$\bar{J}_{ij}(\mathbf{q}) = \frac{1}{\pi}\mathrm{Im}\int^{\epsilon_F} d\epsilon \int_{\mathrm{B.Z.}} d\mathbf{k}\,\mathrm{Tr}[\mathbf{V}_{\mathbf{k},i}\mathbf{G}_{\mathbf{k},ij}\mathbf{V}_{\mathbf{k+q},j}\mathbf{G}_{\mathbf{k+q},ji}], \quad (2)$$

where $\bar{J}$ represents the tensorial form of the non-collinear exchange interactions ($\widehat{H}_{spin} = -\mathbf{S}^i\bar{J}\mathbf{S}^j - K(\mathbf{S}^i\cdot\widehat{z})^2$). The antisymmetric off-diagonal components correspond to DM interactions, while the diagonal components correspond to Heisenberg exchange. In Eq. (2), $\mathbf{G}_{\mathbf{k},ij}$ denotes the Green's function propagator, and $\mathbf{V}_{\mathbf{k},i}$ represents the spin-perturbation potential, which can be obtained from the Kohn-Sham Hamiltonian derived within DFT. These parameters are subsequently Fourier transformed to obtain $J_n^{ij}$ in real space for arbitrary distance exchange pairs, avoiding the need for multiple total-energy calculations.

Finally, while the exchange terms are computed by the MFT, the anisotropy constant $K$ is determined by comparing the DFT total energies of different magnetization directions.

Since we are interested in non-trivial spin textures, of particular interest is the first-neighbor DM interaction ($\vec{D}_1$). Their in-plane components ($|\vec{D}_1^{\mathrm{in}}|$) are represented by the lower-right (triangle) color (see the yellow-green color bar). From Moriya's symmetry rules[51], homostructured 1T TMDCs are expected to have $\vec{D}_{1,2,3} = 0$ due to inversion symmetry. For homostructured 2H phase, $\vec{D}_{1,3}^{\mathrm{in}} = 0$ and $\vec{D}_2 = 0$ due to mirror symmetry whereas $\vec{D}_{1,3}^{\mathrm{out}} \neq 0$ in general. Figure 1c demonstrates that this expected symmetry feature is well captured by our calculations.

In contrast, Janus TMDCs generally host non-zero DM while their mirror plane confines $\vec{D}_2$ along the in-plane direction ($\vec{D}_2^{\mathrm{out}} = 0$). In fact, Fig. 1b shows that $|\vec{D}_1^{\mathrm{in}}|$ are non-zero (greenish in color) in several materials. We note that the magnitude of $|\vec{D}_1^{\mathrm{in}}|$ in some materials is comparable to that of Co/Pt layers (~ 2-3 meV)[52], a prototypical system of strong interfacial DM interactions. Given that the sizable $\vec{D}_1$ can play a crucial role in the formation of topological spin textures, the corresponding materials deserve the careful further investigation (see below).

MFT results of Heisenberg-type spin-spin interactions ($J$) are summarized in Table S2 and Fig. S2. Since antiferromagnetic couplings within triangular lattice can produce magnetic frustration (Fig. S1c), the appearance of blue sectors in Fig. S2 indicates that the corresponding TMDCs can possibly host intriguing spin textures even with small DM[53–55].

It is also noted that longer-range interactions ($J_2$ and $J_3$) can be as large as (or sometimes even larger than) the nearest $J_1$. Importantly, we have considered Heisenberg-type exchange and DM interaction up to the third neighboring pairs (namely, $J_{1,2,3}$ and $\vec{D}_{1,2,3}$). Although MFT provides us even longer couplings ($J_{n\geq4}$ and $\vec{D}_{n\geq4}$) as well, they are an order of magnitude smaller and hence assumed to be negligible in spin texture formations. The capability of computing the longer-range couplings without large supercells is one obvious advantage of using MFT[49,56]. As will be presented below, our simulations in fact reveal that $J_2$ and $J_3$ can be critical in the formation of some spin textures. More detailed information of our MFT results are provided in the Supplementary Information (Table S2).

Finally, we performed atomistic spin simulations to investigate the possible spin textures[57–59]. SPIRIT code was used[60]. To comprehensively explore a large set of possible spin configurations, the simulations were repeated ten times starting from different random initial configurations, each of which runs through $10^6$ steps per material. We also conducted the additional separate spin dynamics simulations based on a different approach; namely, using Landau-Lifshitz-Gilbert (LLG) equation[57] under the same simulation conditions as in the Monte Carlo calculations. This independent simulation can provide the confirmation of our conclusions in regard to the stability of spin superstructures. TMDCs were modeled with the supercells consisting of $100\times100$ magnetic sites, and the results double checked with the $99\times99$ cell to consider the other possible spin structures in a different commensurability (e.g., three-sublattice structures). We also investigated properties such as energy profiles and dynamic responses under external magnetic fields, if needed. Unless otherwise specified, the results presented hereafter refer to zero temperature and zero field conditions.

Among the 54 materials whose spin textures were investigated, half exhibit rather trivial ferromagnetic uniform moment distributions. Here, we focus on the 27 cases, which can be classified into four categories. The first three carry nonzero topological charge: (i) skyrmions, (ii) bimerons, and (iii) maze-like domains; the fourth consists of (iv) exotic spin textures emerging from antiferromagnetic regimes. Materials not discussed in this section include cases with uniform spin polarization or simple domain structures.

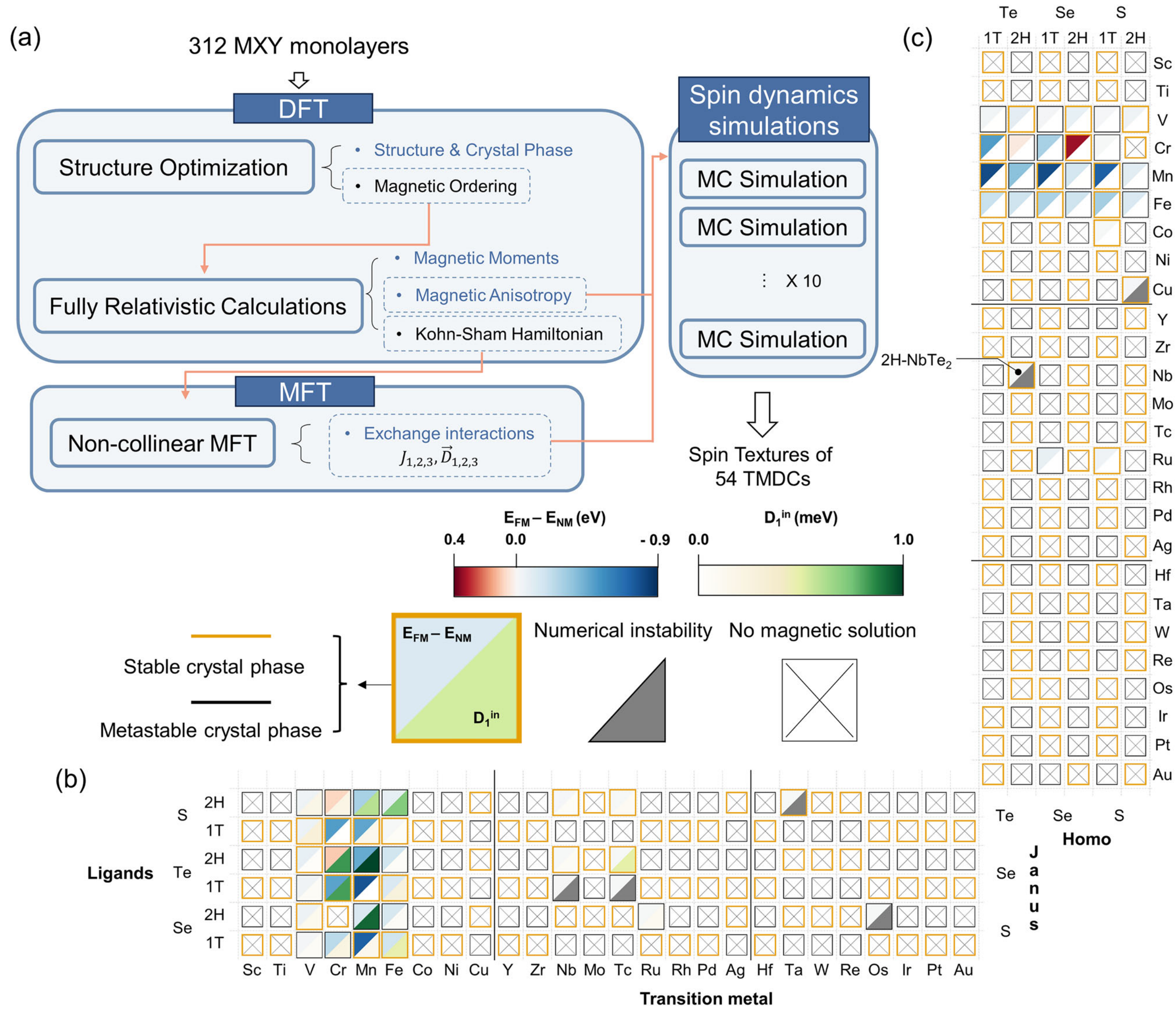

**Fig. 1 | Computational workflow and DFT/MFT screening results for 312 TMDC monolayers.** Suggested workflow (**a**) and calculated DFT and MFT results (**b**, **c**). **a** The workflow consists of DFT calculations followed by MFT and spin-dynamics simulations. The results for Janus- and homo-structured TMDCs are shown in (**b**, **c**), respectively. Each box in (**b**, **c**) represents a specific TMDC system defined by the transition metal, ligand pair, and crystal structure (1T or 2H). The border color of each box indicates the relative structural stability; among the 1T and 2H phases, the orange (black) border denotes the energetically more (less) favorable structure. The upper-left triangular part of each box shows the total energy difference between the ferromagnetic (FM) and non-magnetic (NM) states, $E_{FM} - E_{NM}$ (see the red–blue color bar). Cross symbols (empty or white) indicate materials that do not converge to the FM state. The lower-right triangular part represents the in-plane component of the DM interaction, $D_1^{in}$, ranging from white (nearly zero) to dark green (large values). Six materials (2H-$CuS_2$, 1T-NbSeTe, 2H-$NbTe_2$, 1T-TcSeTe, 2H-TaSTe, and 2H-OsSSe), marked by gray lower-right triangles, correspond to cases in which noncollinear spin DFT calculations fail to reach self-consistency, in contrast to the collinear calculations. Further details of the DFT results, including lattice constants and magnetic moments, are provided in the Supplementary Information (Table S1).

### Skyrmions and skyrmion bundles

Skyrmions are topological quasiparticles theoretically predicted by Tony Skyrme within the framework of nonlinear field theory[61]. Their magnetic realization was proposed by Rößler et al.[62] and experimentally confirmed in chiral magnets[63,64]. Magnetic skyrmions are characterized by continuous swirling of moments and described by a topological charge

$$Q = \frac{1}{4\pi}\int\int d^2\mathbf{r}\cdot\left(\frac{\partial \mathbf{n}}{\partial x}\times\frac{\partial \mathbf{n}}{\partial y}\right), \quad (3)$$

where $\mathbf{n}(\mathbf{r})$ denotes the magnetization unit vector. While $Q = \pm 1$ for isolated skyrmions[65], composite configurations called skyrmion bundles can have different net $Q$[66]. Despite substantial progress, the number of host materials remains limited, and therefore, discovering new systems particularly with high thermal stability and controllability is still a pressing challenge[9,67].

Our workflow predicts four new materials: 2H-RuSSe, 1T-VSSe, 2H-TcSTe, and 1T-FeSTe. See Fig. 2a–d where skyrmion textures are well identified by inverted spin-core contrasts (red/blue) surrounded by Néel-type domains with smoothly rotating moments (visible as white rings or directional arrows). Skyrmion phases in 1T-VSSe and 2H-TcSTe are energetically more stable than the uniform spin polarizations by approximately 0.032 and 0.008 meV/f.u., respectively. In contrast, 2H-RuSSe and 1T-FeSTe host metastable skyrmion structures having, respectively, ~ 0.020 and 0.008 meV/f.u. higher energy than their ferromagnetic ground states.

It is also found that 2H-TcSTe and 1T-FeSTe can host skyrmion bundles: Fig. 2e, f show that smaller skyrmions are embedded within larger ones. Previous studies highlight the unique dynamic properties of 'bundled' textures such as reduced depinning current, enhanced current-driven mobility, and tunable Hall responses[1,5,68,69]. Skyrmion bundles shown in Fig. 2(e) and (f) carry $Q = 0$ and $Q = 4$, respectively, and have slightly higher energy than their conventional skyrmion counterparts by ~ 0.003 and 0.022 meV/f.u., respectively. These relatively small energy differences indicate their possible realization under suitable external conditions. Detailed comparative data are provided in Table S3.

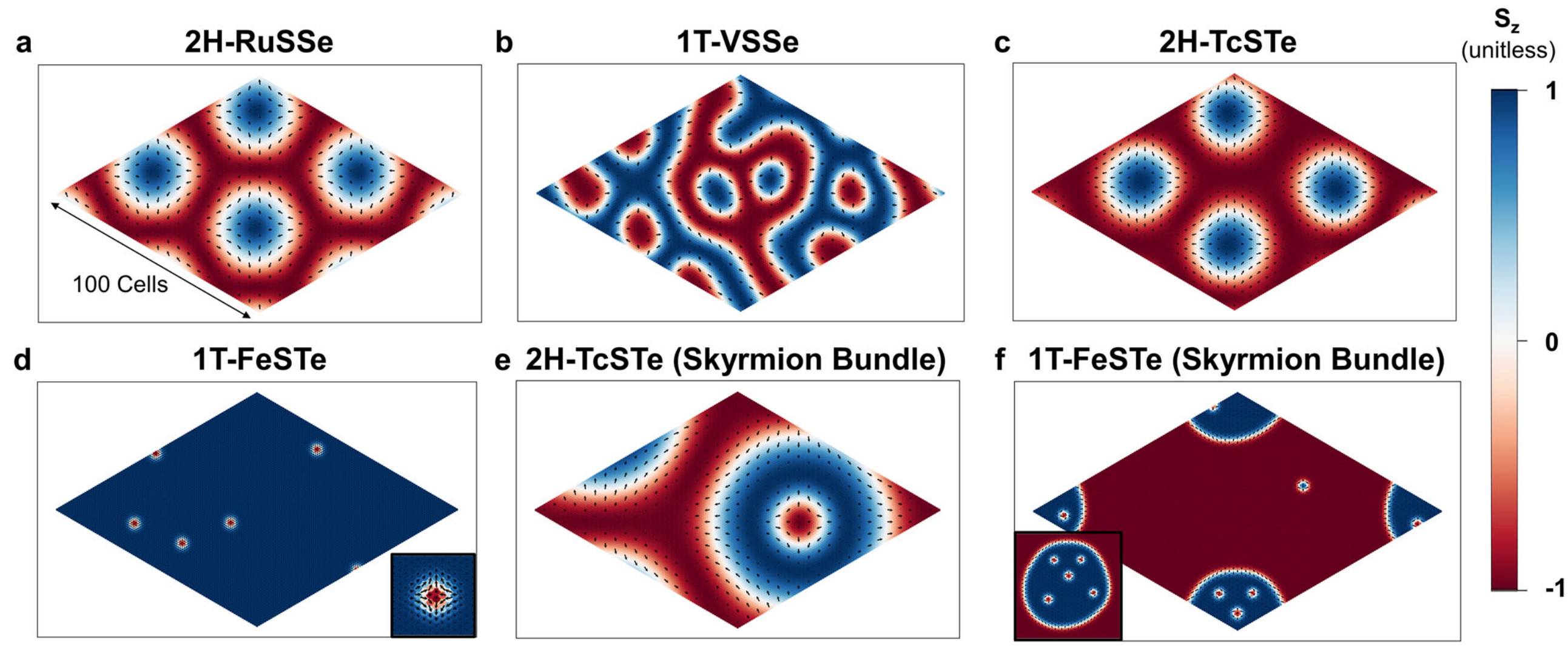

**Fig. 2 | Skyrmion phases in four TMDC monolayers.** Magnetization distributions of the skyrmion phases identified in (**a**) 2H-RuSSe, (**b**) 1T-VSSe, (**c**) 2H-TcSTe, and (**d**) 1T-FeSTe. An enlarged view of a single skyrmion is shown in the inset of (**d**). Skyrmion bundles are also observed in (**e**) 2H-TcSTe and (**f**) 1T-FeSTe. Considering the periodic boundary conditions, an enlarged view of a bundle is shown in the inset of (**f**). Local magnetization directions are visualized using both colors and arrows. The out-of-plane ($z$) component of the magnetization, $S_z$, is represented by the color scale, with blue and red corresponding to upward ($S_z = +1$) and downward ($S_z = -1$) directions, respectively. Black arrows indicate the in-plane components of the magnetization. Each parallelogram in (**a**–**f**) represents the simulated unit cell containing 100 × 100 magnetic ions.

Now we consider the question why skyrmions are stabilized in these four materials. The analysis reveals that they can be categorized into two distinctive cases. In 2H-RuSSe and 2H-TcSTe, DM interaction plays the major role in favoring skyrmion formation (see Table S3). It is well known that, within uniform exchange alignment field, DM term can induce the chiral twist and stabilize skyrmion core by tilting neighboring spins[63,64]. In particular, the in-plane DM components are found to play a more significant role in stabilizing skyrmions than the out-of-plane components, reflecting the presence of Néel-type domain walls around the skyrmion cores and aligning with theoretical expectations for such systems.

In 1T-VSSe and 1T-FeSTe, on the other hand, magnetic frustration (or antiferromagnetic Heisenberg coupling) is more important. Exchange energy gains are 0.237 and 0.008 meV/f.u., respectively, from $J_3 < 0$ whereas DM contributions are clearly smaller (Table S3). By destabilizing uniform spin alignment, antiferromagnetic Heisenberg exchange can promote frustrations and noncollinear spin textures[54]. Similarly, one can also understand the stabilization of skyrmion bundles (see Table S3).

Importantly, it is found that the beyond-nearest-neighbor couplings can be crucial in the workflow. Synthetic simulations only with $J_1$ and $D_1$ (intentional unphysical setting of $J_{2,3} = D_{2,3} = 0$) give rise to the uniform spin texture with no indication of skyrmion for 2H-RuSSe and 1T-VSSe. Also, for 1T-FeSTe, both skyrmions and skyrmion bundles are not stabilized any longer. Further, taking nearest-neighbor coupling model can predict the unphysical stabilization of skyrmions. Skyrmionic phases are found to stabilize in 1T-CrSSe and 1T-CrSeTe within $J_1$–$D_1$ model, which is not the case when $J_{2,3}$ and $D_{2,3}$ are included. Here, despite the strong first-neighbor DM couplings ($|D_1|/J_1$ = 0.15 and 0.29, respectively) and the substantial stabilization energies (0.008 and 0.616 meV/f.u.), skyrmion state becomes unstable due to additional ferromagnetic $J_2$ and $J_3$ (see Table S3). Our results highlight the critical role of often-neglected long-range terms and the importance of their realistic estimations in this kind of searching workflow (Fig. 1a).

## Bimerons

The concept of a meron was introduced as a half-instanton in the context of quark confinement[70]. A pair of merons with identical topological charges ($\pm\frac{1}{2}$, $\pm\frac{1}{2}$) forms a "bimeron", carrying a total topological charge of ± 1 and being analogous to the in-plane-magnetized counterpart of a skyrmion. In this paper, the term "bimeron" broadly encompasses both genuine bimerons and meron–antimeron pairs, unless explicitly differentiated. While the condensed-matter realization of merons or bimerons has been actively explored[71–78], it is generally more challenging than that of skyrmions, highlighting the importance of identifying more stable and flexible material platforms for applications[73,79–81].

Among 312 TMDCs, four compounds are found to host bimerons: 1T-FeSeTe, 1T-CrSeTe, 1T-$RuS_2$, and 2H-$FeS_2$. In Fig. 3, bimerons are clearly visible with the pairs of (blue or red) cores surrounded by vortex-like spin configurations (indicated by black arrows), and can be further verified by computing $Q = \pm\frac{1}{2}$ for each swirling core. More information on their magnetic moments and energetics can be found in Tables S1, S2, and S3.

The formation and stability of bimerons in these four materials can be understood in a manner similar to that of skyrmions. In 1T-FeSeTe and 1T-CrSeTe, DM interactions are the dominant factor: most Heisenberg couplings are ferromagnetic, with the only exception being $J_3 < 0$ in 1T-FeSeTe, and $|\vec{D}_1|$ is sufficiently strong. The ratio $|\vec{D}_1|/J_1$ is 0.28 and 0.29 in 1T-FeSeTe and 1T-CrSeTe, respectively (Fig. S2 and Table S2). This picture is further supported by the analysis of the decomposed energy profiles (Table S3).

On the other hand, $|\vec{D}_1|/J_1$ is small in 2H-$FeS_2$ and 1T-$RuS_2$. By symmetry, $\vec{D}_1^{\text{in}} = 0$, whereas the antiferromagnetic $J_3$ is sizable, with $J_3/J_1 = -0.27$ and $-0.35$, respectively (see Table S2). Thus, in these two materials, bimerons are stabilized mainly by magnetic frustration. On the background of in-plane polarization induced by easy-plane anisotropy ($K < 0$), bimerons emerge due to either strong DM interactions or antiferromagnetic competition.

It is noted that bimerons are not the most stable spin textures in these materials. However, their total energies are only 0.051 and 0.029 meV/f.u.

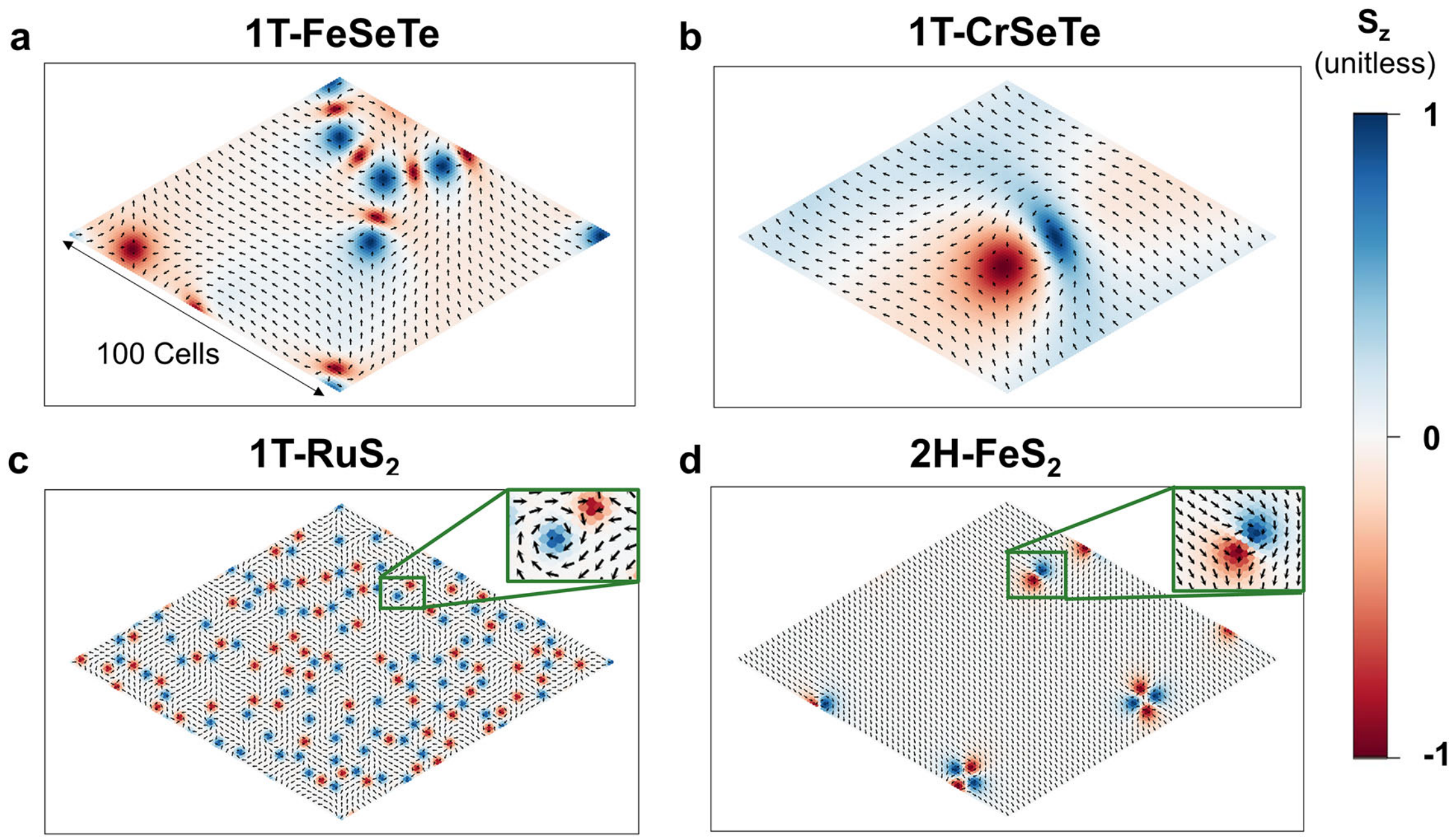

**Fig. 3 | Bimeron phases in four TMDC monolayers.** Magnetization distributions of the bimeron phases identified in (**a**) 1T-FeSeTe, (**b**) 1T-CrSeTe, (**c**) 1T-$RuS_2$, and (**d**) 2H-$FeS_2$. Local magnetization directions are visualized using colors and arrows. The out-of-plane ($z$) component of the magnetization, $S_z$, is represented by the color scale, with blue and red corresponding to upward ($S_z = +1$) and downward ($S_z = -1$) directions, respectively. Black arrows indicate the in-plane components of the magnetization. Insets in (c) and (d) show magnified views highlighting two distinct bimeron core structures and the surrounding swirling spin configurations. Each parallelogram in (**a**–**d**) represents the simulated unit cell containing 100 × 100 magnetic ions.

higher than the uniform ferromagnetic state in 1T-FeSeTe and 1T-CrSeTe, respectively, and 0.163 and 0.672 meV/f.u. higher in 2H-$FeS_2$ and 1T-$RuS_2$ (Table S3). The relatively small energy differences indicate the possible realization under appropriate experimental conditions (e.g., magnetic fields).

### Maze-like domain

In the large-DM limit, circular skyrmions can transform into ramified stripe textures as an attempt to maximize the number of twisted spin pairs and to minimize the total energy. The resulting spin textures have been referred to by several different names in the literature, including "worm-like" domains, "labyrinthine" domains, "stripe skyrmions," or "maze-like" domains[82–84]. Their shapes are seemingly random and hence termed "maze-like," but their stripe width is proportional to the ratio between the exchange and DM interactions[84]. Some maze-like domains are topologically equivalent to skyrmions, and can be transformed into conventional skyrmions under external magnetic fields[85,86]. Their potential applications have been explored in terms of controlling domain formation and motion[87].

Our calculations show that four TMDCs can host maze-like domain structures: 2H-FeSTe, 1T-MnSSe, 1T-$MnS_2$, and 1T-$MnSe_2$. Figure 4a, b present the results for 2H-FeSTe and 1T-$MnS_2$, respectively. Despite the finite unit-cell sizes, the distinct microscopic domain structures are clearly identified.

The stability of maze-like domain structures can be understood in a manner similar to that of skyrmions[85]. In 2H-FeSTe and 1T-MnSSe, we find that a large $|\vec{D}_1|$ plays an important role, and the resulting domains appear as deformed skyrmions. Their total energies are lower than that of the uniform ferromagnetic state by 1.747 and 2.662 meV/f.u., respectively, while the DM contributions amount to 2.480 and 0.647 meV/f.u., respectively (see Table S3).

In the weak-DM regime, maze-like domains can be stabilized through magnetic frustration, particularly via antiferromagnetic $J_2$ or $J_3$ interactions[53,54]. 1T-$MnS_2$ and 1T-$MnSe_2$ represent cases in which symmetry enforces $|\vec{D}_1| = 0$. The energy gains from Heisenberg exchange are 13.639 and 4.761 meV/f.u., respectively, whereas the total energies are 2.779 and 0.519 meV/f.u., respectively, relative to the uniform spin texture (see Table S3). The resulting spin textures are distinct from those induced by DM interactions. As clearly seen for 1T-$MnS_2$ in Fig. 4b, the in-plane spin orientations vary rapidly across each maze wall.

As discussed above, field control of the domain structure is a key feature of maze-like textures. To demonstrate this behavior, we present the field-dependent evolution of the spin structures of 2H-FeSTe (c, e) and 1T-$MnS_2$ (d, f) in Fig. 4c–f. It is noted that the characteristic maze-like domain structures gradually disappear and eventually transform into skyrmions under sufficiently strong magnetic fields.

A crucial distinction lies in their chirality characteristics: DM-induced skyrmions maintain a fixed chirality, whereas frustration-induced skyrmions exhibit variable chirality, enabling the coexistence of skyrmions and antiskyrmions within the same domain. The inset of Fig. 4f illustrates Bloch-type and Néel-type skyrmions, as well as antiskyrmions, coexisting in the same frustration-induced skyrmion lattice system[53,54,88].

### Spin textures in antiferromagnets

Intriguing spin textures can also be realized in antiferromagnetic materials ($J_1 < 0$), offering useful advantages such as field immunity, ultrafast dynamics, and enhanced stability[89–91]. While this possibility has been studied mainly in bulk materials, new TMDC candidates can provide an attractive alternative platform[92–94]. For convenience, we classify 15 materials with $J_1 < 0$ into four cases, considering the relative strengths of $J_{1,2,3}$[95]: (i) ferromagnetic, (ii) Néel phase (triangular coplanar), (iii) row-wise antiferromagnetic (RW-AFM), and (iv) spin spiral, as presented in Fig. 5.

Eleven materials are found in the regime of Néel-type order (Fig. 5). Significant in-plane DM interactions can lead to formation of skyrmions or spin spirals commensurate with the Néel order. Conversely, out-of-plane DM interactions act as an effective anisotropy, enhancing easy-plane alignment[96]. In the materials investigated here, only vortex and helical spin structures are observed largely due to the weak in-plane DM interactions. Fig. S3a, b display the vortex and spin spiral results for 2H-$CrSe_2$ and 1T-$CrTe_2$, respectively. While out-of-plane DM interaction can stabilize in-plane vortices, this effect is insufficient to make the vortex state the ground state (Table S3). These states frequently persist depending on the initial conditions in our simulations, suggesting their experimental accessibility. In materials such as 1T-MnSTe and 1T-VSeTe, where DM interactions are present, topological texture might more likely be realized. Although antiferromagnetic Néel-type skyrmions were previously suggested[95], it was not observed in our simulations.

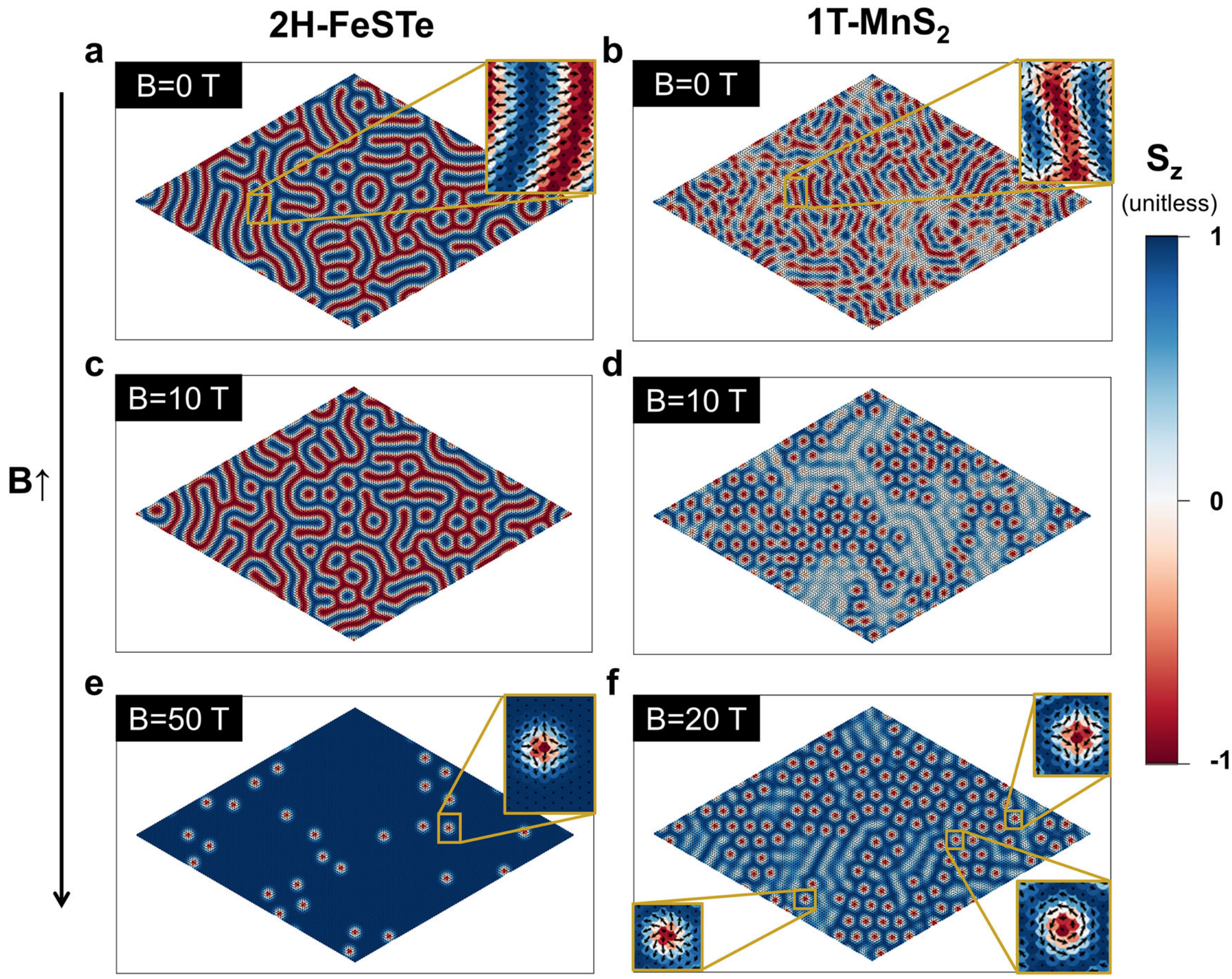

**Fig. 4 | Maze-like domains in 2H-FeSTe and 1T-$MnS_2$, and their evolution with varying magnetic field strength.** Local spin directions are represented by the color scale and black arrows. The color scale denotes the out-of-plane component of the magnetization, $S_z$. **a**, **b** Magnetization patterns of 2H-FeSTe and 1T-$MnS_2$, respectively, at zero magnetic field ($B$ = 0 T), exhibiting characteristic maze-like domain structures. Insets show close-up views of representative regions. **c**, **d** Evolution of these domains at an intermediate magnetic field of $B$ = 10 T, where the patterns begin to change under the influence of the external field. **e**, **f** Domain structures at higher magnetic fields ($B$ = 50 and 20 T, respectively), where the maze-like domains transform into skyrmions. The figure highlights two distinct types of maze-like domains: in 2H-FeSTe, the domains can be viewed as deformations of skyrmions stabilized by nearest-neighbor DM interactions, whereas in 1T-$MnS_2$, the maze-like domains originate from magnetic frustration, characterized by rapid variations of the in-plane spin components across each maze wall.

2H-MnSSe and 1T-$CrS_2$ locate in the RW-AFM regime (Fig. 5). 1T-CrSe2 retains the bare RW-AFM pattern due to $|\vec{D}| = 0$. In 2H-MnSSe, on the other hand, $|\vec{D}_{1,2,3}| = 0.95, 0.33$, and 0.27 meV, respectively, with $|J_1|$ = 6.95 meV, leading to a spin spiral stabilized on top of the RW-AFM background (see Fig. S3c). In this spin texture, the spiral exhibits a stripe pattern, with one spiral rotating from left to right and the next from right to left. The DM contribution to this stabilization is 1.204 meV/f.u. with respect to the uniform RW-AFM order (Table S3). While these states can be converted into skyrmion structures, we find that it requires very high field of ~ 350 T.

2H-$MnTe_2$ is located near the boundary between the RW-AFM and Néel phases (Fig. 5). Indeed, we find a domain structure comprising both phases (see Fig. S3d); the RW-AFM domains display spin-spiral configurations analogous to those in 2H-MnSSe. To our knowledge, this type of composite spin texture has not been well explored, and its properties and potential applications therefore constitute an interesting direction for future research.

Other spin textures discussed in the literature (e.g., inter-chained skyrmions) are not observed, presumably because $D_3$ is not sufficiently large[95].

## Discussion

Before closing, we briefly discuss the limitations of the present computational approach together with the experimental context in which the predicted candidates should be interpreted. While the PBE functional can overestimate magnetic moment formation trends [97,98], it has nevertheless proven reliable for itinerant magnetic systems of the type considered here. In particular, PBE-level DFT has been shown to reproduce experimentally observed spin-spiral periods, skyrmion-lattice length scales, and interfacial DM interaction strengths in prototypical systems[52,99,100]. The subsequent MFT-based extraction of $J$ and $D$ parameters depends on the choice of localized basis used to project the Kohn–Sham Hamiltonian and is therefore quantitatively sensitive to this choice. The implementation adopted here has, however, been benchmarked against alternative choices across representative magnetic systems[56], and the MFT is by now well established for first-principles evaluation of magnetic interactions, including DM interaction[101]. Finally, atomistic spin simulations based on classical spins neglect quantum fluctuations and are subject to finite-size effects. Despite these caveats, this approach has been shown to quantitatively reproduce experimentally observed skyrmion phases and transition behavior,

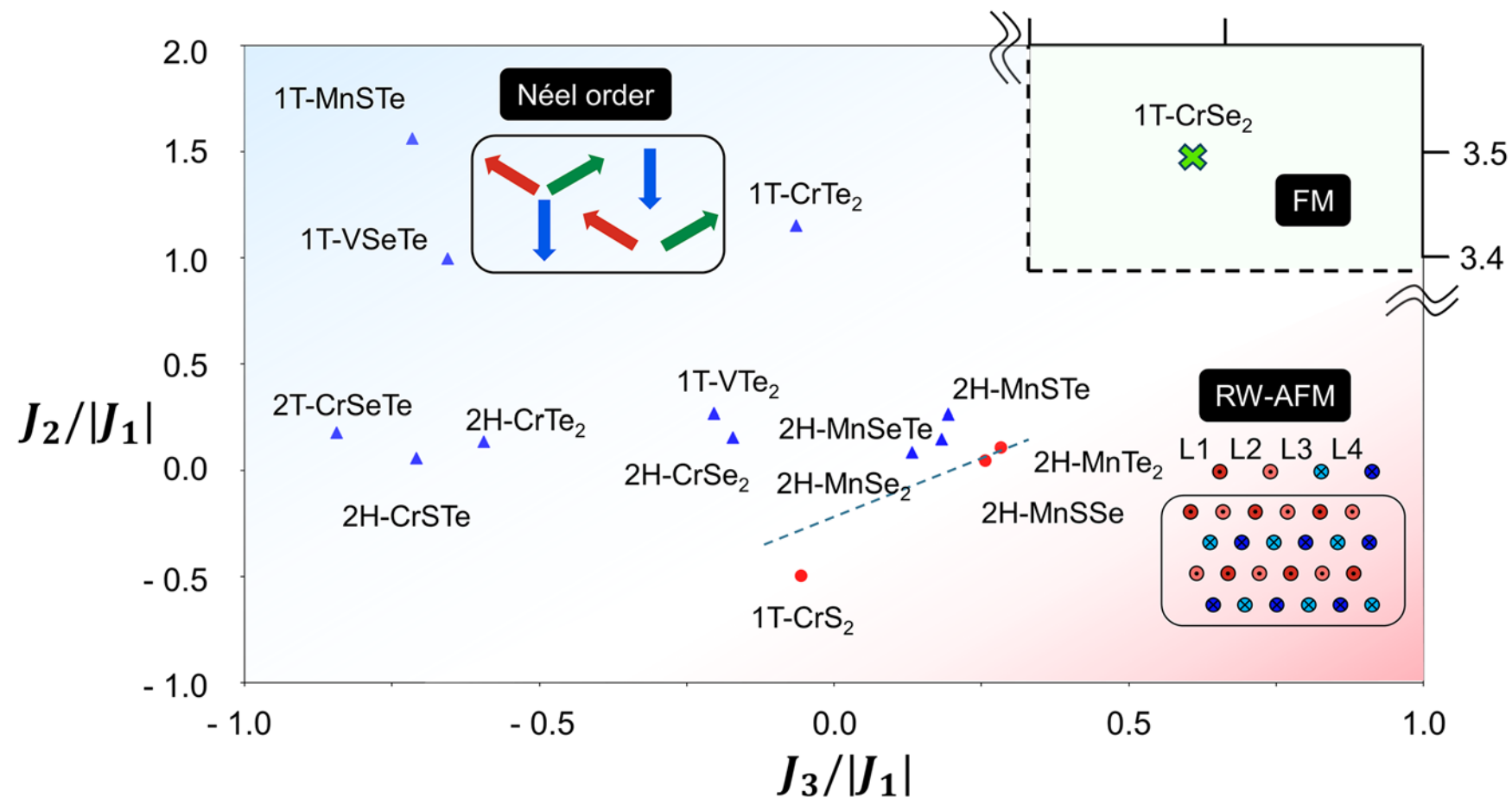

**Fig. 5 | Phase diagram of 15 materials with antiferromagnetic $J_1$ and representative spin textures.** Phase diagram classified in terms of $J_2/|J_1|$ and $J_3/|J_1|$, showing the regions corresponding to the Néel (blue triangles), row-wise antiferromagnetic (RW-AFM, red circles), and ferromagnetic (FM, green cross) phases. Insets illustrate schematic spin configurations of the Néel and RW-AFM phases.

supporting the reliability of the present workflow for identifying candidate materials[102–105].

Our main simulations are carried out at zero temperature, while thermal fluctuations destroy the underlying long-range magnetic order above $T_c$ and, even below $T_c$, can disrupt the predicted spin textures themselves by overcoming the energy barrier protecting them. While the formation-energy difference $\Delta E$ that we report identifies which textures are energetically preferred at zero temperature, the relevant quantity for thermal stability is the topological collapse barrier $E_b$ together with the attempt-frequency asymmetry between the metastable texture and the trivially ordered background. Indeed, theoretical calculations and experimental observations confirm that metastable spin textures with small $\Delta E$ can persist on macroscopic timescales, in some cases up to room temperature [106–108].

On the materials side, we note that many of the homo-chalcogen compounds relevant to our candidate set (both the homo candidates themselves and the parent compounds of the Janus candidates) have already been realized even at the mono- or few-layer level[109–113]. On the Janus side, monolayer synthesis was first demonstrated for MoSSe by $H_2$-plasma stripping followed by thermal selenization[29,114] and has since been extended through low-energy Se-ion pulsed-laser deposition[115] and, more recently, through room-temperature atomic-layer substitution of Au-tape-exfoliated parent dichalcogenides yielding mm-scale single-crystal Janus WSSe monolayers[116]. In particular, selenization of $VTe_2$ has produced monolayer Janus 1T-VSeTe – one of the materials introduced in the current study – providing direct experimental support for the feasibility of our predicted systems[117].

In this work, we performed a systematic first-principles-based investigation of magnetic interactions and emergent spin textures in TMDC monolayers, including both homo-structured and Janus variants. By combining high-throughput DFT calculations, MFT, and large-scale atomistic spin-dynamics simulations, we established a physically transparent and reproducible workflow that connects crystal structure, exchange interactions, and emergent spin textures at the materials level. Using this workflow, we screened a broad materials space and identified a diverse set of magnetic TMDCs hosting nontrivial spin textures. The principal candidates include 2H-RuSSe, 1T-FeSTe, 2H-TcSTe, and 1T-VSSe (skyrmions); 1T-FeSeTe, 2H-$FeS_2$, 1T-$RuS_2$, and 1T-CrSeTe (bimerons); 2H-FeSTe, 1T-MnSSe, 1T-$MnS_2$, and 1T-$MnSe_2$ (maze-like domains). Our results demonstrate that skyrmions, bimerons, maze-like domains, and antiferromagnetic vortex or spiral textures arise from distinct microscopic mechanisms governed by the competition between Heisenberg exchange interactions, DM interactions, and magnetic frustration. We further showed that these textures exhibit characteristic responses to external magnetic fields, including field-driven transformations from maze-like domains to skyrmion phases. Taken together, this study establishes TMDC monolayers as a versatile platform for noncollinear magnetism and provides an interaction-based screening framework for understanding and designing magnetic textures in low-dimensional materials.

## Methods

### DFT calculations

The DFT calculations including structural optimization were conducted by using Vienna ab initio Simulation Package (VASP)[118]. The convergence criteria for the energy and the Hellmann-Feynman force were set to $10^{-7}$ eV and 0.001 eV/Å, respectively. To simulate monolayer geometry, a 15 Å vacuum layer was taken into account. Plane-wave basis cutoff was determined after the convergence test for each material. The corresponding criterion was 0.001 eV, and the typical cutoff energy 500 eV. Perdew-Burke-Ernzerhof (PBE) functional of generalized gradient approximation (GGA)[97] was chosen to approximate the exchange-correlation energy. $\Gamma$-centered **k** mesh was adopted and the number of grid determined from the convergence test. The typical mesh is $17 \times 17 \times 1$. In order to construct the full-band spin-dependent Hamiltonian with local indices for the next MFT procedure, we additionally performed DFT calculations with OpenMX software package[48,119,120]. Non-collinear DFT calculations incorporating spin-orbit coupling were performed using fully relativistic $j$-dependent pseudopotentials to describe the magnetic anisotropy energy and anisotropic exchange interactions. The same **k** grid with VASP calculation and 1000 Ry energy cutoff were used.

### Magnetic force theory

'Jx' code was used for MFT calculations[121]. The expression of Eq.(2) in the manuscript comes through

$$[\mathbf{G}_{\mathbf{k},ij}]^{\sigma}_{l_1 l_2} = \sum_n \frac{|w^{\sigma}_{l_1,\mathbf{k},i}\rangle\langle w^{\sigma}_{l_2,\mathbf{k},j}|}{z - \epsilon_{\sigma,n,\mathbf{k}} + i0^+}, \quad (4)$$

where $\sigma = \uparrow, \downarrow$ are spin indices, and the orbital $l_1$ and $l_2$ belong to the atomic site $i$ and $j$, respectively. $|w^{\sigma}_{l_1,\mathbf{k},i}\rangle$ refers to the local state. The final real-space value is obtained through Fourier transformation,

$$J(\mathbf{R}_{ij}) = \sum_{l_1 l_2} \tilde{J}^{l_1 l_2}(\mathbf{R}_{ij}) = \sum_{l_1 l_2} \sum_{\mathbf{q}} \tilde{J}^{l_1 l_2}_{ij}(\mathbf{q})\, e^{-i\mathbf{R}_{ij}\cdot\mathbf{q}}. \quad (5)$$

We use the same number of **q** grids as with DFT calculations.

## Data availability

The datasets generated and/or analyzed during the current study are not publicly available due to the large aggregate size of the raw VASP, OpenMX,

Jx, and SPIRIT output files but are available from the corresponding author on reasonable request. No custom code was developed for this study; all calculations were performed using publicly available software packages (VASP, OpenMX, Jx, and SPIRIT) cited in the Methods section.

## Code availability

No custom code was developed for this study. All calculations were performed using publicly available software packages (VASP, OpenMX, Jx, and SPIRIT) cited in the Methods section.

## Acknowledgements

This work was supported by the National Research Foundation of Korea (NRF) grant funded by the Korea government (MSIT) (Nos. RS-2023-00253716, RS-2025-0559042, RS-2025-02243032).

## Author contributions

Y.C. performed DFT and MFT calculations and conducted atomistic spin-dynamics simulations together with D.H.K. Y.C. and M.J.H. wrote the manuscript with contributions from D.H.K. M.J.H. supervised the project. All authors have read and approved the manuscript.

## Competing interests

The authors declare no competing interests.

## Additional information

**Supplementary information** The online version contains supplementary material available at https://doi.org/10.1038/s41524-026-02162-2.

**Correspondence** and requests for materials should be addressed to Myung Joon Han.